\documentclass[journal, twocolumn]{IEEEtran}

\usepackage[utf8]{inputenc} 
\usepackage[english]{babel}
\usepackage[colorlinks=true]{hyperref}
\usepackage{amsthm,amsfonts,amssymb,amsmath, microtype}
\usepackage[colorlinks=true]{hyperref}

\usepackage{xcolor}
\usepackage{graphicx}
\usepackage{mathtools}

\graphicspath{{fig/}}

\newtheorem{theorem}{Theorem}

\newtheorem{lemma}{Lemma}

\newtheorem{definition}{Definition}

\newcommand{\etal}[1]{et al.}
\newcommand{\vct}[1]{\mathbf{#1}}
\newcommand{\mtx}[1]{\mathbf{#1}}

\newcommand{\OA}{\text{OA}}

\newcommand{\OAstrength}{t}
\newcommand{\OAlevels}{\sigma}
\newcommand{\OAfactors}{n} 
\newcommand{\OAsymbols}{\mathbb{Z}}
\newcommand{\OAruns}{M}
\newcommand{\OAindex}{\lambda}

\makeatletter 
 \hypersetup{pdftitle = {Derandomizing compressed sensing with combinatorial design},
       pdfauthor = {Peter Jung, Richard Kueng,Dustin G.\ Mixon},
       pdfsubject = {Compressed sensing},
       pdfkeywords = {}
      }
\makeatother

\begin{document}

\title{Derandomizing compressed sensing with combinatorial design}

\author{%
  \IEEEauthorblockN{Peter Jung\IEEEauthorrefmark{1}, Richard Kueng\IEEEauthorrefmark{2},
    Dustin G.\ Mixon\IEEEauthorrefmark{3}}\\
  \IEEEauthorblockA{
 \IEEEauthorblockA{\IEEEauthorrefmark{1}\IEEEauthorblockA{Communications and Information Theory Group, Technische Universit\"{a}t Berlin, Germany}}\\
\IEEEauthorrefmark{2}\IEEEauthorblockA{
	Department of Computing + Mathematical Sciences \& Institute for Quantum Information and Matter, California Institute of Technology, USA
}}\\
 \IEEEauthorblockA{\IEEEauthorrefmark{2}\IEEEauthorblockA{Department of Mathematics, Ohio State University, USA}}
}

\maketitle

\begin{abstract}
Compressed sensing is the art of reconstructing structured
$n$-dimensional vectors from substantially fewer measurements than
naively anticipated. A plethora of analytic reconstruction guarantees
support this credo. The strongest among them are based on deep results
from large-dimensional probability theory that require a considerable
amount of randomness in the measurement design. Here, we demonstrate
that derandomization techniques allow for considerably reducing the
amount of randomness that is required for such proof strategies. More,
precisely we establish uniform s-sparse reconstruction guarantees for
$C s \log (n)$ measurements that are chosen independently from
strength-four orthogonal arrays and maximal sets of mutually unbiased
bases, respectively. These are highly structured families of
$\tilde{C} n^2$ vectors that imitate signed Bernoulli and standard
Gaussian vectors in a (partially) derandomized fashion.
\end{abstract}


\section{Introduction and main results}

\subsection{Motivation}

Compressed sensing is the art of reconstructing structured signals from substantially fewer measurements than would naively be required for standard techniques like least squares. 
Although not entirely novel, rigorous treatments of this observation \cite{donoho_compressed_2006, candes_robust_2006} spurred considerable scientific attention from 2006 on, see e.g.\ \cite{foucart_mathematical_2013,eldar_compressed_2012} and references therein.
While deterministic results do exist, the strongest theoretic convergence guarantees still rely on randomness. Broadly, these can be grouped into two families:
\begin{enumerate}
\item \emph{generic measurements} such as independent Gaussian, or Bernoulli vectors. Such an abundance of randomness allows for establishing very strong results by following comparatively simple and instructive proof techniques. 
The downside is that concrete implementations do require a lot of randomness. In fact, they might be too random to be useful for certain applications.
\item \emph{structured measurements} such as random rows of a Fourier, or Hadamard matrix. In contrast to generic measurements, these feature a lot of structure that is geared towards applications. Moreover, sampling random rows from a fixed matrix does require very little randomness. E.g.\ $\log (n)$ random bits are required to sample a random DFT row while an i.i.d.\ Bernoulli vector consumes $n$ bits of randomness. 
Structure and comparatively little randomness have a downside, however. Theoretic convergence guarantees tend to be weaker than their generic counterparts. It should also not come as a surprise that the necessary proof techniques become considerably more involved.
\end{enumerate}
Typically, results of type 1) precede results of type 2). 
Phase retrieval via PhaseLift is a concrete example for such a development. Generic convergence guarantees \cite{candes_phaselift_2013,candes_solving_2012} preceded (partially) de-randomized results \cite{gross_partial_2015,kueng_low_2017}.
Compressed sensing is special in this regard. The two seminal works \cite{donoho_compressed_2006,candes_robust_2006} from 2006 provided both results almost simultaneously.
This had an interesting consequence. Despite considerable effort, to this date there still seems to be a gap between both proof techniques. 

Here, we try to close this gap by applying a method that is very well established in theoretical computer science: \emph{partial derandomization}.
We start with a proof technique of type 1) and considerably limit the amount of randomness required for it to work. While doing so, we keep careful track of the ``amount of randomness'' that is still necessary. Finally, we replace the original (generic) random measurements with pseudo-random ones that mimic them in a sufficiently accurate fashion. 
Our results highlight that this technique \emph{almost} allows for bridging the gap between existing proof techniques for generic and structured measurements: the results are still strong, but require slightly more randomness than choosing vectors uniformly from a bounded orthogonal system, such as Fourier or Hadamard vectors. 

There is a also a didactic angle to this work: within the realm of signal processing, partial-derandomization techniques have been successfully applied to matrix reconstruction \cite{kueng_low_2017,kabanava_stable_2016} and phase retrieval via PhaseLift \cite{gross_partial_2015,kueng_spherical_2015,kueng_low_2016}. 
Although similar in spirit, the more involved nature of these problems may obscure the key ideas, intuition and tricks behind such an approach. 
However, the same techniques have not yet been applied to the original problem of compressed sensing. Here, we fill this gap and, in doing so, provide an introduction to partial derandomization techniques by example. To preserve this didactic angle, we try to keep the presentation as simple and self-contained as possible. 

Finally, one may argue that compressed sensing has not fully lived up to the high expectations of the community yet, see e.g.\ \cite{tropp_mathematical_2017}.
Arguably, one of the most glaring problems for applications is the requirement of choosing individual measurements at random\footnote{Existing deterministic constructions, see e.g.\ \cite{bandeira_road_2013}, do not (yet) yield comparable statements.}.
While we are not able to fully overcome this drawback here, the methods described in this work do limit the amount of randomness required to generate individual structured measurements. We believe that this may help to reduce the discrepancy between ``what can be proved'' and ``what can be done'' in a variety of concrete applications.

\subsection{Preliminaries on compressed sensing}

Compressed sensing aims at reconstructing $s$-sparse vectors $\vct{x} \in \mathbb{C}^n$  from $m \ll n$ linear measurements:
\begin{equation*}
\vct{y} = \mtx{A} \vct{x} \in \mathbb{C}^m.
\end{equation*}
Since $m \ll n$, the matrix $\mtx{A}$ is singular and there are infinitely many solutions to this equation. A convex penalizing function is used to promote sparsity among these solutions. Typically, this penalizing function is the $\ell_1$-norm $\| \vct{z} \|_{\ell_1} = \sum_{i=1}^n |z_i |$:
\begin{align}
\underset{\vct{z} \in \mathbb{C}^n}{\textrm{minimize}} & \quad \| \vct{z} \|_{\ell_1} \label{eq:l1minimization}\\
\textrm{subject to} & \quad \mtx{A} \vct{z} = \vct{y}
 \nonumber
\end{align}
Mathematical proofs for convergence to the correct solution $\vct{x} \in \mathbb{C}^n$ have been established for different measurement matrices $\mtx{A}$. By and large, they require  randomness in the sense that each row $\vct{a}_i \in \mathbb{C}^n$ of $\mtx{A}$ is an \emph{independent} copy of a random vector $\vct{a} \in \mathbb{R}^n$. Prominent examples include
\begin{enumerate}
\item $m = C s \log (n/s)$ standard complex Gaussian measurements: $\vct{a}_{g}\sim \mathcal{N} (\vct{0}, \mathbb{I}/\sqrt{2})+i\mathcal{N} (\vct{0}, \mathbb{I}/\sqrt{2})$, 
\item $m = C s \log (n/s)$ signed Bernoulli (Rademacher) measurements: $\vct{a}_{sb} \sim \left\{ \pm 1 \right\}^n$,
\item $m = C s \log^4 (n)$ random rows of a DFT matrix: $\vct{a}_{f} \sim \left\{ \vct{f}_1,\ldots,\vct{f}_n \right\}$,
\item for $n=2^d$: $m = C s \log^4 (n)$ random rows of a Hadamard matrix: $\vct{a}_{h} \sim \left\{ \vct{h}_1,\ldots,\vct{h}_n \right\}$.
\end{enumerate}
A rigorous treatment of all these cases can be found in Ref.~\cite{foucart_mathematical_2013}.
Here, and throughout this work, $C>0$ denotes an absolute constant whose exact value depends on the context, but it is always independent of the problem parameters $n,s$ and $m$.
It is instructive to compare the amount of randomness that is required to generate one instance of the random vectors in question. A random signed Bernoulli vector $\vct{a}_{sb} \in \mathbb{R}^n$ requires $n$ random bits (one for each coordinate), while a total of $d = \log_2 (n)$ random bits suffice to select a random row $\vct{a}_{h} \in \mathbb{R}^n$ of a Hadamard matrix. 
A comparison between complex standard Gaussian vectors $\vct{a}_{g} \in \mathbb{C}^n$ and random Fourier vectors $\vct{a}_{f} \in \mathbb{C}^n$ indicates a similar discrepancy. In summary: highly structured random vectors, like $\vct{a}_{f},\vct{a}_{h}$ require exponentially fewer random bits to generate than generic random vectors, like $\vct{a}_g, \vct{a}_{sb}$. 
Importantly, this transition from generic measurements to highly structured ones comes at a price. The number of measurements required in case (1) and (4) scales poly-logarithmically in $n$. More sophisticated approaches allow for converting this offset into a polylogarithmic scaling in $s$ rather than $n$
 \cite{rudelson_sparse_2008,cheraghchi_restricted_2013}.
Another, arguably even higher price, is hidden in the proof techniques behind these results. They are considerably more involved.

The following two subsections are devoted to introduce formalisms that allow for partially de-randomizing signed Bernoulli vectors and complex standard Gaussian vectors, respectively. 

\subsection{Partially de-randomizing signed Bernoulli vectors}

Throughout this work, we endow $\mathbb{C}^n$ with the standard inner product $\langle \vct{x},\vct{y} \rangle = \sum_{i=1}^n \bar{x}_i y_i$. We denote the associated (Euclidean) norm by $\| \vct{z} \|_{\ell_2}^2 = \langle \vct{z},\vct{z} \rangle$.
Let $\vct{a}_{sb} = \sum_{i=1}^n \epsilon_i \vct{e}_i$ be a signed Bernoulli vector with coefficients $\epsilon_i\sim \left\{ \pm 1 \right\}$ chosen independently at random (Rademacher random variables). Then,
\begin{equation}
\mathbb{E} \left[\epsilon_i \bar{\epsilon}_j \right]=\mathbb{E} \left[ \epsilon_i \epsilon_j \right]  = \delta_{ij}
\label{eq:pairwise_independence}
\end{equation}
which is equivalent to demanding
\begin{align}
\mathbb{E} \left[ \langle \vct{y}, \vct{a}_{sb} \rangle \langle\vct{a}_{sb},\vct{z} \rangle \right]
=& \sum_{i,j=1}^n \mathbb{E} \left[ \epsilon_i \bar{\epsilon}_j \right] \bar{y}_i z_j = \langle \vct{y},\vct{z} \rangle\quad \forall \vct{y},\vct{z}.
\label{eq:isotropy_stronger}
\end{align}
Independent sign entries are sufficient, but not necessary for this feature.
Indeed, suppose that $n=2^d$ is a power of two. Then the rows of a Sylvester Hadamard matrix  $\vct{h}_1,\ldots,\vct{h}_n$ correspond to a particular subset of sign vectors. 
Let $\vct{a}_h \in \mathbb{R}^n$ be the random vector arising from choosing a Hadamard row uniformly at random. Then,
\begin{align*}
\mathbb{E} \left[ \langle \vct{y}, \vct{a}_h \rangle \langle \vct{a}_h,\vct{z} \rangle \right]
= \frac{1}{n} \sum_{i=1}^n \langle \vct{y},\vct{h}_i \rangle \langle \vct{h}_i, \vct{z} \rangle
= \langle \vct{y},\vct{z} \rangle \quad \forall \vct{y},\vct{z},
\end{align*}
because the Hadamard rows $\vct{h}_i$'s are proportional to an orthonormal basis and have norm $\sqrt{n}$. 
This in turn implies that the coordinates $h_i,h_j \in \left\{ \pm 1 \right\}$ of a randomly selected Hadamard matrix row obey \eqref{eq:pairwise_independence}, despite not being independent instances of random signs.  
This feature is called \emph{pairwise independence} and naturally generalizes to $k \geq 2$:

\begin{definition}[$k$-wise independence] \label{def:k_wise_independence}
Fix $k \geq 2$ and let $\epsilon_i$ denote independent instances of a signed Bernoulli random variable.
We call a random sign vector $\vct{a}\in\left\{\pm1 \right\}^n$ $k$-wise independent, if its components $a_1,\ldots,a_n$ obey
\begin{equation*}
\mathbb{E} \left[ \prod_{i=1}^k a_{i_k} \right] = \mathbb{E} \left[ \prod_{i=1}^k \epsilon_{i_k} \right]
\end{equation*}
for all $k$-tuples of indices $1 \leq i_1,\ldots,i_k \leq n$.
\end{definition}

Explicit constructions for $k$-wise independent vectors are known for any $k$ and $n$.
In this work we focus on particular constructions that rely on generalizing the following instructive example. Fix $n=4$ and consider the rows of the following matrix:
\begin{equation*}
\left(
\begin{array}{rrrr}
  1 &  1 & -1 & -1 \\
  1 & -1 &  1 & -1 \\
  1 & -1 & -1 &  1
\end{array}
\right)
\end{equation*}
The first two rows summarize all possible length-two combinations of $\pm 1$. The coefficients of the third row correspond to their entry-wise product. Hence, it is completely characterized by the first two. The three row vectors are \emph{not} mutually independent. 
Nonetheless, each subset of two rows does mimic independent behavior: 
all possible length-two combinations of $\pm 1$ occur exactly once. 
This ensures that a randomly selected row is pairwise independent in the sense that its coefficients obey Eq.~\eqref{eq:pairwise_independence}.

This simple example may readily be generalized.
 A binary $M \times n$ \emph{orthogonal array} of \emph{strength} $t$ is a sign matrix  $\mtx{O} \in \left\{\pm 1 \right\}^{M \times n}$ such that every selection of $t$ rows contains all elements of $\left\{ \pm 1 \right\}^{t}$ an equal number of times. 

Several different explicit constructions of orthogonal arrays are known.
A simple counting argument reveals that the number of rows must obey $M \geq O(n^{t/2})$. This number scales polynomially in the array strength $t$ -- a potentially exponential improvement over the ``full'' array that lists all $2^n$ possible elements of $\left\{ \pm 1 \right\}^n$. In turn, selecting a random row of $\mtx{O}$ only requires $\log_2 (M) \geq \mathcal{O}( t \log_2 (n))$ random bits
and produces a random vector that is $t$-wise independent according to Definition~\ref{def:k_wise_independence}. We refer to Sec.~\ref{sec:numerics} and Ref.~\cite{Hedayat:OAbook} for a more thorough treatment of this concept.

\subsection{Partially derandomizing complex standard Gaussian vectors}

Let us now discuss another general purpose tool for (partial) de-randomization. 
Concentration of measure implies that $n$-dimensional standard complex Gaussian vectors concentrate sharply around the complex sphere $\sqrt{n} \mathbb{S}^{n-1}$ of radius $\sqrt{n}$.
Hence, they behave very similarly to vectors $\vct{a}_{s} \in \mathbb{C}^n$ chosen uniformly from this sphere. Such random vectors obey the following formula for any $k \in \mathbb{N}$ and any $\vct{z} \in \mathbb{C}^n$:
\begin{align*}
\mathbb{E} \left[ | \langle \vct{z},\vct{a}_s \rangle|^{2k} \right]
=& n^k \int_{\vct{w} \in \mathbb{S}^{n-1}} | \langle \vct{z},\vct{w} \rangle|^{2k} \mathrm{d}\vct{w}\\
 =& n^k \binom{n+k-1}{k}^{-1} \| \vct{z} \|_{\ell_2}^{2k}.
\end{align*}
Here, $\mathrm{d}\vct{w}$ denotes the uniform measure on the complex unit sphere $\mathbb{S}^{n-1} \subset \mathbb{C}^n$.
This formula characterizes even moments of this uniform distribution\footnote{For comparison, a complex standard Gaussian vector obeys $\mathbb{E} \left[ | \langle \vct{z},\vct{a}_g \rangle |^{2k} \right] = k! \| \vct{z} \|_{\ell_2}^{2k}$ instead.}.
The concept of $k$-designs \cite{delsarte_spherical_1991} uses this moment formula as a starting point for partial de-randomization. 
Roughly speaking, a $t$-design is a finite subset of $\sqrt{n}$-length vectors such that the uniform distribution over these vectors reproduces the uniform measure on $\sqrt{n} \mathbb{S}^{n-1}$ up to $k$-th moments. More precisely:

\begin{definition} \label{def:design}
A set of $N$ vectors $\left\{ \vct{w}_i \right\}_{i=1}^n \subset \sqrt{n} \mathbb{S}^{n-1}$ with length $\sqrt{n}$ is called a \emph{(complex projective) $t$-design} if a randomly chosen vector $\vct{a}_{(t)}$ obeys for any $1 \leq k \leq t$
\begin{equation*}
\mathbb{E} \left[ | \langle \vct{z},\vct{a}_{(t)} \rangle|^{2k} \right]
= n^k \binom{n+k-1}{k}^{-1} \| \vct{z} \|_{\ell_2}^{2k}
\quad \forall\vct{z} \in \mathbb{C}^n.
\end{equation*}
\end{definition}

(Spherical) $t$-designs were originally developed as cubature formulas for the real-valued unit sphere \cite{delsarte_spherical_1991}. The concept has since been extended to other sets. A generalization to the complex projective space $\mathbb{C}P^{n-1}$ gives rise to Definition~\ref{def:design}.
Complex projective $t$-designs are known to exist for any $t$ and any dimension $n$, see e.g.\ \cite{bajnok1992construction,korevaar_chebyshev_1994, hayashi2005reexamination}. 
However, explicit constructions for $t \geq 3$ are notoriously difficult to find. 
In contrast, several explicit families of 2-designs have been identified. 
Here, we will focus on one such family.
Two orthonormal bases $\left\{ \vct{b}_i \right\}_{i=1}^n$ and $\left\{ \vct{c}_i \right\}_{i=1}^n$ of $\mathbb{C}^n$ are called \emph{mutually unbiased}
 if
\begin{equation}
\left| \langle \vct{b}_i, \vct{c}_j \rangle \right|^2 = \frac{1}{n}
\quad \textrm{for all} \quad i,j \in \left[n\right]=\left\{1,\ldots,n \right\}. \label{eq:mutually_unbiased}
\end{equation}
A prominent example for such a basis pair are the standard basis and the Fourier, or Hadamard, basis, respectively. 
One can show that at most $n+1$ different orthonormal bases exist that have this property in a pairwise fashion \cite[Theorem~3.5]{bandyopadhyay_new_2002}. Such a set of $n+1$ bases is called a \emph{maximal set of mutually unbiased bases} (MMUB). 
For instance, in $n=2$ the standard basis together with
\begin{equation*}
\frac{1}{\sqrt{2}}\left(
\begin{array}{c}
1 \\
1
\end{array}
\right),
\frac{1}{\sqrt{2}}\left(
\begin{array}{c}
1 \\
-1
\end{array}
\right),
\frac{1}{\sqrt{2}}\left(
\begin{array}{c}
1 \\
i
\end{array}
\right), \\
\frac{1}{\sqrt{2}}\left(
\begin{array}{c}
1 \\
-i
\end{array}
\right)
\end{equation*}
forms a MMUB. Importantly, MMUBs are always (proportional to) 2-designs \cite{klappenecker_mutually_2005}.
Explicit constructions exist for any prime power dimension $n$ and one can  ensure that the standard basis is always one of them. Here we point out one construction that is particularly simple if the dimension is (an odd) prime $n \geq 5$ \cite{klappenecker_constructions_2004}:
The standard basis vectors $\vct{e}_1,\ldots,\vct{e}_n \in
\mathbb{C}^n$ together with all vectors whose entry-wise coefficients
correspond to
\begin{equation}
\left[ \vct{b}_{\alpha,\lambda} \right]_k = \frac{1}{\sqrt{n}} \omega_n^{(k+\alpha)^3 + \lambda (k+\alpha)}  \label{eq:mub}
\end{equation}
form a MMUB. Here $\omega_n = \exp \left( \frac{2\pi i}{n} \right)$ is a $n$-th root of unity.
The parameter $\alpha \in \left[n\right]$ singles out one of the $n$ different bases, while $\lambda \in \left[n\right]$ labels the $n$ corresponding basis vectors.
Excluding the standard basis, this set of $n^2$ vectors corresponds to all time-frequency shifts of a discrete Alltop sequence $\left[\vct{f}\right]_k = \omega_n^{k^3}$ \cite{alltop_complex_1980}.

\subsection{Main results}

\begin{theorem}[CS from orthogonal array measurements] \label{thm:oa4}
Suppose that a matrix $\mtx{A}$ contains $m \geq C s \log (2n)$ rows that are chosen independently from an orthogonal array with strength four. Then, with probability at least $1- 2\mathrm{e}^{-\tilde{c} m}$,  any $s$-sparse $\vct{x} \in \mathbb{C}^n$ can be recovered from $\vct{y} = \mtx{A} \vct{x}$ by means of algorithm~\eqref{eq:l1minimization}.
\end{theorem}

\begin{theorem}[CS from time-frequency shifted Alltop sequences] \label{thm:mub}
Let $n \geq 5$ be  prime and
suppose that $\mtx{A}$ contains $m \geq C s \log (2n)$ rows that correspond to random time-frequency shifts of the Alltop sequence \eqref{eq:mub} in dimension $n$. Then, with probability at least $1- \mathrm{e}^{-\tilde{c} m}$,  any $s$-sparse $\vct{x} \in \mathbb{R}^n$ can be recovered from $\vct{y} = \mtx{A} \vct{x}$ by means of algorithm~\eqref{eq:l1minimization}.
\end{theorem}

This result actually generalizes to measurements that are sampled from a maximal set of mutually unbiased bases (excluding the standard basis). Time-frequency shifts of the Alltop sequence are one concrete construction that applies to prime dimensions only.

Note that the cardinality of all Alltop shifts is $n^2$. Hence, $2 \log_2 (n)$ random bits suffice to select a random time-frequency shift. 
In turn, a total of
\begin{equation}
 2\log_2 (n) m \simeq 2C s \log^2 (n) \label{eq:main_random_bits}
\end{equation}
random bits are required for sampling a complete measurement matrix $\mtx{A}$.
This number is exponentially smaller than the number of random bits required to generate a matrix with independent complex Gaussian entries.
A similar comparison holds true for random signed Bernoulli matrices and columns sampled from a strength-4 orthogonal array.

Highly structured families of vectors -- such as rows of a Fourier, or Hadamard matrix -- require even less randomness to sample from: only $\log_2 (n)$ bits are required to select such a row uniformly at random.
However, existing convergence guarantees are weaker than the main results presented here. They require an order of $Cs  \mathrm{polylog}(s) \log (n)$ random measurements to establish comparable results. Thus, the total number of random bits required for such a procedure scales like $Cs \mathrm{polylog}(s)\log^2(n)$.
Eq.~\eqref{eq:main_random_bits} still establishes a logarithmic improvement in terms of sparsity.

The recovery guarantees in Theorem~\ref{thm:oa4} and \ref{thm:mub} can be readily extended to ensure stability with respect to noise corruption in the measurements and robustness with respect to violations of the model assumption of sparsity. We refer to Sec.~\ref{sec:stable} for details.

We also emphasize that there are results in the literature that establish compressed sensing guarantees comparable, or even less, randomness. 
Obviously, deterministic constructions are the extreme case in this regard. Early results suffer from a ``quadratic bottleneck''. The number of measurements must scale quadratically in the sparsity: $m \simeq s^2$. Although this obstacle was overcome, existing progress is still comparatively mild. Refs.~\cite{bourgain_explicit_2011,mixon_explicit_2015,bandeira_conditional_2017} establish deterministic convergence guarantees for $m \simeq s^{2-\epsilon}$, where $\epsilon>0$ is a (very) small constant. 

Closer in spirit to this work is Ref.~\cite{bandeira_derandomizing_2016}.
There, the authors employ the Legendre symbol -- which is well known for its pseudorandom behavior -- to partially derandomize a signed Bernoulli matrix. In doing so, they establish uniform $s$-sparse recovery from $m \geq C s \log^2 (s) \log (n)$ measurements that require an order of $s \log (s) \log (n)$ random bits to generate. 
Compared to the main results presented here, this result gets by with less randomness, but requires more measurements. The proof technique is also very different.

To this date, the strongest de-randomized reconstruction guarantees hail from a close connection between $s$-sparse recovery and Johnson-Lindenstrauss embeddings \cite{baraniuk_simple_2008,krahmer_new_2011}. These have a wide range of applications in modern data science. Kane and Nelson \cite{kane_derandomized_2010} established a very strong partial de-randomization for such embeddings. This result may be used to establish uniform $s$-sparse recovery for $m = C s \log (n/s)$ measurements that require an order of $s \log \left(s \log (n/s)\log (n/s) \right)$ random bits. 
This result surpasses the main results presented here in both sampling rate and randomness required. 

However, this strong result follows from ``reducing'' the problem of $s$-sparse recovery to a (seemingly) very different problem: find Johnson-Lindenstrauss embeddings. Such a reduction typically does not preserve problem-specific structure. In contrast, the approach presented addresses the problem of sparse recovery directly and relies on tools from signal processing. In doing so, we maintain structural properties that are common in several applications of $s$-sparse recovery.
Orthogonal array measurements, for instance, have $\pm 1$-entries. 
This is well-suited for the single pixel camera \cite{duarte_single_2008}. 
Alltop sequence constructions, on the other hand, have successfully been applied to stylized radar problems \cite{herman_resolution_2009}. 
Both types of measurements also have the property that every entry has unit modulus. 
This is an important feature for the application of CDMA \cite{tropp_cdma_2003}.
Having pointed out these high level connections, we want to emphasize that careful, problem specific adaptations may be required to rigorously exploit these.
The framework developed here may serve as a guideline on how to achieve this goal in concrete scenarios.

\section{Proofs}

\subsection{Textbook-worthy proof for real-valued compressed sensing with Gaussian measurements}

This section is devoted to summarizing an elegant argument that is originally due to Rudelson and Vershynin \cite{rudelson_sparse_2008}, see also \cite{kabanava_analysis_2015, dirksen_gap_2016, tropp_convex_2015} for arguments that are similar in spirit.
This argument only applies to $s$-sparse recovery of real-valued signals. We will generalize a similar idea to the complex case later on.

In this work we are concerned with \emph{uniform} reconstruction guarantees: With high probability a single realization of the measurement matrix $\mtx{A}$ allows for reconstructing \emph{any} $s$-sparse vector $\vct{x}$ by means of $\ell_1$-regularization \eqref{eq:l1minimization}.
A necessary pre-requisite for uniform recovery is the demand that no $s$-sparse vector is contained in the kernel, or \emph{nullspace}, of $\mtx{A}$. This condition is captured by the \emph{nullspace property} (NSP).
Define 
\begin{equation}
T_s =  \left\{ \vct{z} \in \mathbb{S}^{n-1}:\; \| \vct{z} \|_{\ell_1} \geq 2 \sigma_s (\vct{z} ) \right\} \subset \mathbb{S}^{n-1},
\label{eq:Ts}
\end{equation}
where $\sigma_s (\vct{x})= \inf_{\| \vct{z} \|_0 \leq s} \| \vct{x} - \vct{z} \|_{\ell_1} \quad \textrm{for} \quad \vct{x} \in \mathbb{C}^n$
is the approximation error (measured in $\ell_1$-norm) one incurs when approximating $\vct{x}$ with a $s$-sparse vector.
A matrix $\mtx{A}$ obeys the NSP of order $s$ if
\begin{equation}
\inf_{\vct{z} \in T_s} \| \mtx{A} \vct{z} \|_{\ell_2} >0. \label{eq:nsp}
\end{equation}
The set $T_s$ is a subset of the unit sphere that contains all normalized $s$-sparse vectors. This justifies the informal definition of the NSP: no $s$-sparse vector is an element of the nullspace of $\mtx{A}$. 
Importantly, the NSP is not only necessary, but also sufficient for uniform recovery, see e.g.\ \cite[Theorem~4.5]{foucart_mathematical_2013}. Hence, universal recovery of $s$-sparse signals readily follows from establishing Rel.~\eqref{eq:nsp}. 
The nullspace property and its relation to $s$-sparse recovery has long been somewhat folklore. We refer to Ref.~\cite{foucart_mathematical_2013} for a discussion of its origin.

The following powerful statement allows for exploiting generic randomness in order to establish nullspace properties. It is originally due to Gordon \cite{gordon_milmans_1988}, but we utilize a more modern reformulation, see \cite[Theorem 9.21]{foucart_mathematical_2013}.

\begin{theorem}[Gordon's escape through a mesh] \label{thm:gordon}
Let $\mtx{A} \in \mathbb{M}_{m \times n}$ be a real-valued standard Gaussian matrix and let $E \subseteq \mathbb{S}^n$ be a subset of the real-valued unit sphere. Define the \emph{Gaussian width}
$
\ell (E) = \mathbb{E} \sup_{\vct{z} \in E} \langle \vct{a}_g, \vct{z} \rangle,
$
where the expectation is over realizations $\vct{a}_g \sim \mathcal{N}(0,\mathbb{I})$ of a standard Gaussian random vector.
Then, for $t \geq 0$ the bound
\begin{equation*}
\inf_{\vct{z} \in E} \| \mtx{A} \vct{z} \|_{\ell_2} \geq \sqrt{m-1} - \ell (E) - t
\end{equation*}
is true with probability at least $1 - \mathrm{e}^{-t^2/2}$.
\end{theorem}

This is a deep statement that connects random matrix theory to geometry: the Gaussian width is a rough measure of the size of the set $E \subseteq \mathbb{S}^n$. 
Setting $E=T_s$ allows us to conclude that a matrix $\mtx{A}$ encompassing $m$ independent Gaussian measurements is very likely to obey the $s$-NSP \eqref{eq:nsp}, provided that $m-1$ exceeds $\ell(T_s)^2$. 
In order to derive an upper bound on $\ell (T_s)$, we may use the following inclusion
\begin{equation*}
T_s \subset 2 \mathrm{conv} \left( \Sigma_n^s \right),
\end{equation*}
see e.g.\ \cite[Lemma~3]{kabanava_analysis_2015} and \cite[Lemma~4.5]{rudelson_sparse_2008}.
Here, $\Sigma^n_s \subseteq \mathbb{S}^n$ denotes the set of all $s$-sparse vectors with unit length. In turn,
\begin{equation}
\ell (T_s ) \leq 2 \mathbb{E} \sup_{\vct{z} \in \mathrm{conv} (\Sigma^s_n)} \langle \vct{a}_g, \vct{z} \rangle 
= 2 \mathbb{E} \sup_{\vct{z} \in \Sigma^n_s} \langle \vct{a}_g, \vct{z} \rangle, \label{eq:gaussian_process}
\end{equation}
because the linear function $\vct{z} \mapsto \langle \vct{a}_g, \vct{z} \rangle$ achieves its maximum value at the boundary $\Sigma_s^n$ of the convex set $\mathrm{conv} \left( \Sigma_s^n \right)$.
The right hand side of \eqref{eq:gaussian_process} is the expected supremum of a Gaussian process indexed by $\vct{z} \in \Sigma_s^n$.  \emph{Dudley's inequality} \cite{dudley_sizes_1967}, see also \cite[Theorem~8.23]{foucart_mathematical_2013}, states
\begin{align*}
\mathbb{E} \sup_{\vct{z} \in \Sigma^n_s} \langle \vct{a}_g, \vct{z} \rangle \leq 4 \sqrt{2} \int_0^1 \sqrt{\ln \left(\mathcal{N}\left(\Sigma_s^n, \| \cdot \|_{\ell_2}, u \right) \right)}, \mathrm{d}u
\end{align*}
where $\mathcal{N}(\Sigma_s^n, \| \cdot \|_{\ell_2}, u )$ are covering numbers associated with the set $\Sigma_s^n$. They are defined as the smallest cardinality of a $u$-covering net with respect to the Euclidean distance. 
A volumetric counting argument yields $\mathcal{N}(\Sigma_s^n, \| \cdot \|_{\ell_2}, u ) \leq \left(\frac{\mathrm{e}n}{s} \right)^s\left(1+\frac{2}{u} \right)^s$
and Dudley's inequality therefore implies
\begin{equation*}
\ell (T_s) \leq c \sqrt{s \log \left( \mathrm{e}n/s\right)},
\end{equation*}
where $c$ is an absolute constant. 
This readily yields the following assertion.

\begin{theorem}[NSP for Gaussian measurements]
A number of $m \geq c s \log (\mathrm{e}n/s)$ independent real-valued Gaussian measurements obeys the (real-valued) $s$-NSP with high probability at least $1-\mathrm{e}^{-\tilde{c}m}$.
\end{theorem}

This argument is exemplary for generic proof techniques:
strong results from probability theory allow for  establishing close-to-optimal results in a relatively succinct fashion.

\subsection{Extending the scope to subgaussian measurements}

The extended arguments presented here are largely due to Dirksen, Lecue and Rauhut \cite{dirksen_gap_2016}. Again, we will focus on the real-valued case.

Gordon's escape through a mesh is only valid for Gaussian random matrices $\mtx{A}$.
Novel methods are required to extend this proof technique beyond this idealized case. 
Comparatively recently, Mendelson provided one by  generalizing Gordon's escape through a mesh \cite{mendelson_learning_2015,koltchinskii_bounding_2015}.

\begin{theorem}[Mendelson's small ball method, Tropp's formulation \cite{tropp_convex_2015}] \label{thm:mendelson}
Suppose that $\mtx{A}$ is a random $m \times n$ matrix whose rows correspond to $m$ independent realizations of a random vector $\vct{a} \in \mathbb{R}^n$. 
Fix a set $E \subseteq \mathbb{R}^n$,  and define
\begin{align*}
Q_\xi (\vct{a},E)=& \inf_{\vct{z} \in E} \mathrm{Pr} \left[ \left| \langle \vct{z},\vct{a} \rangle \right| \geq \xi \right] \quad \textrm{for} \quad \xi >0,\\
W_m (\vct{a},E) =& \mathbb{E} \sup_{\vct{z} \in E} \langle \vct{z}, \vct{h} \rangle
\quad \textrm{where} \quad \vct{h} = \frac{1}{\sqrt{m}} \sum_{i=1}^m \epsilon_i \vct{a}_i \in \mathbb{R}^n ,
\end{align*}
is the empirical average over $m$ independent copies  of $\vct{a}$ weighted by uniformly random signs $\epsilon_i \sim \left\{ \pm 1 \right\}$.
Then, for any $t, \xi >0$ 
\begin{equation*}
\inf_{\vct{z} \in E} \| \mtx{A} \vct{z} \|_{\ell_2} \geq \xi \sqrt{m} Q_{2\xi} (\vct{a},E) - 2W_m (\vct{a},E) - \xi t\label{eq:mendelson}
\end{equation*}
with probability at least $1-2\mathrm{e}^{-t^2/2}$.
\end{theorem}

It is worthwhile to point out that for real-valued Gaussian vectors this result recovers Theorem~\ref{thm:gordon} up to constants. Fix $\xi >0$ of appropriate size. Then, $E \subseteq \mathbb{S}^n$ ensures that $\xi Q_{2\xi} (\vct{a}_g,E)$ is constant.
Moreover, $W_m (\vct{a}_g,E)$ reduces to the usual Gaussian width $\ell (E)$.

Mendelson's small ball method can be used to establish the nullspace property  for independent random measurements $\vct{a} \in \mathbb{R}^n$ that exhibit \emph{subgaussian} behavior:
\begin{equation}
\mathbb{E} \exp \left(\theta \langle \vct{y},\vct{a} \rangle \right)
\leq \exp \left( \frac{\theta^2}{2}\| \vct{y} \|_{\ell_2}^2 \right)
\quad \textrm{for all} \quad \vct{y} \in \mathbb{R}^n,\;\theta >0.
\label{eq:subgaussian}
\end{equation}
Signed Bernoulli vectors are a concrete example: $\left[ \vct{a}\right]_k = \epsilon_k$ is an independent instance of a Rademacher random variable.
Signed Bernoulli vectors obey
\begin{align}
\mathbb{E} \left[ \langle \vct{z},\vct{a}_{\mathrm{sb}} \rangle^2 \right]
	=& \sum_{i,j=1}^n \mathbb{E} \left[ \epsilon_i \epsilon_j \right] z_i z_j = \| \vct{z} \|_{\ell_2}^2 \quad\forall  \vct{z} \in \mathbb{R}^n. \label{eq:isotropy}
\end{align}
Direct computation also reveals
\begin{align}
\mathbb{E} \left[ \langle \vct{z}, \vct{a}_{\mathrm{sb}} \rangle^4 \right]
=& \sum_{i,j,k,l=1}^n \mathbb{E} \left[ \epsilon_i \epsilon_j \epsilon_k \epsilon_l \right] z_i z_j z_k z_l \nonumber  \\
=& \sum_{i=1}^n \mathbb{E} \left[ \epsilon_i^4 \right] z_i^4 +3 \sum_{i \neq j} \mathbb{E} \left[ \epsilon_i^2 \right] \mathbb{E} \left[ \epsilon_j^2 \right] z_i^2 z_j^2 \nonumber \\
=& \sum_{i=1}^n z_i^4 + 3 \sum_{i \neq j} z_i^2 z_j^2 = 3 \| \vct{z} \|_{\ell_2}^4 - 2 \| \vct{z} \|_{\ell_4}^4 \nonumber \\
\leq & 3 \| \vct{z} \|_{\ell_2}^4, \label{eq:4th_moment}
\end{align}
because there are 3 possible pairings of four indices.

Now, set $E = T_s \subset \mathbb{S}^n$.

An application of the Paley-Zygmund inequality then allows for bounding the parameter $Q_{2\xi} (\vct{a}_{\mathrm{sb}}, T_s)$ in Mendelson's small ball method from below:
\begin{align*}
Q_{2\xi} (\vct{a}_{\mathrm{sb}}, T_s )
\geq & \inf_{\vct{z} \in \mathbb{S}^n} \mathrm{Pr} \left[ |  \langle \vct{z},\vct{a}_{sb} \rangle| \geq 2\xi \right] \\
=& \inf_{\vct{z} \in \mathbb{S}^n} \mathrm{Pr} \left[ \langle \vct{z},\vct{a}_{sb}\rangle^2 \geq 4\xi^2 \mathbb{E} \left[ \langle \vct{z},\vct{a}_{sb} \rangle^2 \right] \right] \\
\geq & \inf_{\vct{z} \in \mathbb{S}^n} \left( 1- 4\xi^2 \right)^2 \frac{ \mathbb{E} \left[ \langle \vct{z}, \vct{a}_{sb} \rangle^2 \right]^2}{\mathbb{E} \left[ \langle \vct{z},\vct{a}_{sb} \rangle^4 \right]} 
\geq  \frac{(1-4\xi^2)^2}{3}.
\end{align*}
This lower bound is constant for any $\xi \in (0,1)$.

Next, note that $X_{\vct{z}} = \langle \vct{z},\vct{h} \rangle$ is a stochastic process that is indexed by $\vct{z} \in \mathbb{R}^n$. This process is centered ($\mathbb{E} X _{\vct{z}}=0$) and Eq.~\eqref{eq:subgaussian} implies that it is also subguassian (at least for any $\vct{z} \in \Sigma_s^n$).
Moreover, $\mathbb{E} \left[ | X_\vct{z} - X_{\vct{y}}|^2 \right]^{1/2} = \| \vct{z}-\vct{y} \|_{\ell_2}^2$ readily follows from \eqref{eq:isotropy}. 
Unlike Gordon's escape through a mesh, Dudley's inequality does remain valid for such stochastic processes with subgaussian marginals.
We can now repeat the width analysis from the previous section to obtain
\begin{align*}
W_m (\vct{a}_{sb}, T_s) \leq 2 \mathbb{E} \sup_{\vct{z} \in \Sigma_s^n} \langle \vct{z}, \vct{h} \rangle \leq c \sqrt{s \log (\mathrm{e}n/s)}.
\end{align*}
Fixing $\xi >0$ sufficiently small, setting $t =\tilde{c} \sqrt{m}$ and inserting these bounds into Eq.~\eqref{eq:mendelson} yields the following result.

\begin{theorem}[NSP for signed Bernoulli measurements]
A matrix $\mtx{A}$ encompassing $m \geq C s \log (\mathrm{e}n/s)$ random signed Bernoulli measurements obeys the real-valued $s$-NSP with probability at least $1- \mathrm{e}^{\tilde{c}m}$.
\end{theorem}

A similar result remains valid for other classes of independent measurements with subgaussian marginals \eqref{eq:subgaussian}.

\subsection{Generalization to complex-valued signals and partial de-randomization}

The nullspace property, as well as its connection to uniform $s$-sparse recovery readily generalizes to complex-valued $s$-sparse vectors.
A similar extension applies to Mendelson's small ball method:

\begin{theorem}[Mendelson's small ball method for complex vector spaces]
\label{thm:mendelson_complex}
Suppose that the rows of $\mtx{A}$ correspond to $m$ independent copies of a random vector $\vct{a} \in \mathbb{C}^n$.
Fix a set $E \subset \mathbb{C}^n$ and define
\begin{align*}
Q_\xi (\vct{a},E) =& \inf_{\vct{z} \in E} \mathrm{Pr} \left[ | \langle \vct{z},\vct{a} \rangle| \geq \xi \right]\quad \textrm{for} \quad \xi >0, \\
W_m (\vct{a},E) =& \mathbb{E} \sup_{\vct{z} \in E} \left| \langle \vct{z},\vct{h} \rangle \right| \quad \textrm{where} \quad \vct{h} = \frac{1}{\sqrt{m}} \sum_{i=1}^m \epsilon_i \vct{a}_i .
\end{align*}
Then, for any $t,\xi >0$
\begin{align*}
\inf_{\vct{z} \in E} \| \mtx{A} \vct{z} \|_{\ell_2} \geq \sqrt{2} \left( \xi \sqrt{m} Q_{2^{3/2} \xi}/2 - 2 W_m (E,\vct{a}) -  \xi t \right)
\end{align*}
with probability at least $1-2 \mathrm{e}^{-t^2/2}$.
\end{theorem}

Such a generalization was conjectured by Tropp \cite{tropp_convex_2015}, but we are not aware of any rigorous proof in the literature. We provide one in Subsection~\ref{sub:mendelson_proof} and believe that such an extension may be of independent interest. This extension allows for generalizing the arguments from the previous subsection to the complex-valued case.

Let us now turn to the main scope of this work: partial de-randomization. 
Effectively, Mendelson's small ball method reduces the task of establishing nullspace properties to bounding the two parameters $Q_{2^{3/2}\xi} (\vct{a},T_s)$ and $W_m (\vct{a},T_s)$ in an appropriate fashion. 
A lower bound on the former readily follows from the Paley-Zygmund inequality, provided that the random vector $\vct{a}$ obeys
\begin{align*}
\mathbb{E} \left[ | \langle \vct{a},\vct{z}\rangle|^2 \right] =& \| \vct{z} \|_{\ell_2}^2\quad \textrm{for all} \quad \vct{z} \in \mathbb{C}^n& \textrm{(isotropy)},  \\
\mathbb{E} \left[ | \langle \vct{a},\vct{z} \rangle |^4 \right] \leq &C_4 \| \vct{z} \|_{\ell_2}^4& \textrm{(4h moment bound)},
\end{align*}
where $C_4 >0$ is a constant:
\begin{equation}
Q_{2^{3/2}\xi} (\vct{a},T_s)
\geq C_4^{-1} \left(1-8\xi^2 \right)^2\quad \textrm{for any} \quad \xi >0.
\label{eq:Q}
\end{equation}
In contrast, establishing an upper bound on $W_m (\vct{a},T_s)$ via Dudley's inequality requires subgaussian marginals \eqref{eq:subgaussian} (that must not depend on the ambient dimension). This implicitly imposes stringent constraints on \emph{all} moments simultaneously.
An additional assumption allows to considerably weaken these demands:
\begin{align}
\max_{1 \leq k \leq n} | \langle \vct{e}_k,\vct{a} \rangle|^2=&1 \quad \textrm{almost surely} & \textrm{(incoherence)}.
\end{align}

Incoherence has long been identified as a key ingredient for developing $s$-sparse recovery guarantees. Here, we utilize it to establish an upper bound on $W_m (\vct{A},T_s)$ that does not rely on subgaussian marginals.

\begin{lemma} \label{lem:W}
Let $\vct{a} \in \mathbb{C}^n$ be a random vector that is isotropic and incoherent. Let $T_s \subset \mathbb{C}^n$ be the complex-valued generalization of the set defined in Eq.~\eqref{eq:Ts} and assume $m \geq \log (2n)$. Then,
\begin{equation}
W_m (\vct{a},T_s) \leq 4 \sqrt{2 s\log (2n)}. \label{eq:W}
\end{equation}
\end{lemma}

This bound only requires an appropriate scaling of the first two moments (isotropy). 
However, this partial derandomization comes at a price: the bound scales logarithmically in $n$ rather than $n/s$.
We defer a proof of this statement to Subsection~\ref{sub:W} below.
Inserting the bounds \eqref{eq:Q} and \eqref{eq:W} into the assertion of Theorem~\ref{thm:mendelson_complex} readily yields the main technical result of this work:

\begin{theorem} \label{thm:derandomization}
Suppose that $\vct{a} \in \mathbb{C}^n$ is a random vector that obeys incoherence, isotropy and the 4th moment bound. 
Then, choosing
\begin{equation*}
m \geq C s \log (n)
\end{equation*}
instances of $\vct{a}$ uniformly at random results in a measurement matrix $\mtx{A}$ that obeys the complex-valued nullspace property of order $s$ with probability at least $1-2 \mathrm{e}^{-\tilde{c}m}$.
\end{theorem}

In complete analogy to the real-valued case, the complex nullspace property ensures uniform recovery of $s$-sparse vectors $\vct{x} \in \mathbb{C}^n$ from linear measurements of the form $\vct{y} = \mtx{A} \vct{x}$ via algorithm \eqref{eq:l1minimization}.

\subsection{Recovery guarantee for strength-four orthogonal arrays}

Suppose that $\vct{a}_{oa} \in \left\{ \pm 1 \right\}^n$ is chosen uniformly from an orthogonal array with strength 4. By definition
\begin{equation*}
\| \vct{a}_{oa} \|_{\ell_\infty} = | \pm 1 | = 1,
\end{equation*}
which establishes \emph{incoherence}. Moreover, the components $a_i$ of $\vct{a}_{oa}$ obey $\mathbb{E} \left[ a_i a_j \right] = \mathbb{E} \left[ \epsilon_i \epsilon_j \right]= \delta_{ij}$, because 4-wise independence necessarily implies 2-wise independence. Isotropy readily follows:
\begin{align*}
\mathbb{E} \left[ | \langle \vct{z},\vct{a} \rangle|^2 \right]
= \sum_{i,j} \mathbb{E} \left[ a_i \bar{a}_j \right] \bar{z}_i z_j = \langle \vct{z},\vct{z} \rangle \quad \forall \vct{z} \in \mathbb{C}^n.
\end{align*}
Finally, 4-wise independence suffices to establish the 4th moment bound.
By assumption $\mathbb{E} \left[ a_i a_j \bar{a}_k \bar{a}_l \right] = \mathbb{E} \left[ \epsilon_i \epsilon_j \epsilon_k \epsilon_l \right]$
and we may thus infer
\begin{align*}
& \mathbb{E} \left[ |\langle \vct{z}, \vct{a}_{oa} \rangle|^4 \right] = \sum_{i,j,k,l=1}^n \mathbb{E} \left[ \epsilon_i \epsilon_j \epsilon_k \epsilon_l \right] \bar{z}_i \bar{z}_j z_k z_l \\
=& \sum_{i=1}^n \mathbb{E} \left[ \epsilon_i^4 \right] |z_i|^4 \\
 +& \sum_{i \neq j} \mathbb{E} \left[ \epsilon_i^2 \right] \mathbb{E} \left[ \epsilon_j^2 \right]
\left( \sum_{i \neq j} \bar{z}_i^2 z_j^2 + 2 \sum_{i \neq j} |z_i|^2 \sum_j |z_j|^2 \right)  \\
\leq & \| \vct{z} \|_{\ell_4}^4 + 3 \|\vct{z} \|_{\ell_2}^4 \leq 4 \| \vct{z} \|_{\ell_2}^4.
\end{align*}
Therefore $\vct{a}_{oa}$ meets all the requirements of Theorem~\ref{thm:derandomization}. The first main result then readily follows from the fact that the complex nullspace property ensures uniform recovery of all $s$-sparse signals.

\subsection{Recovery guarantee for mutually unbiased bases}

Suppose that $\vct{a}_{mub} \in \mathbb{C}^n$ is chosen uniformly from a maximal set of $n$ mutually unbiased bases (excluding the standard basis) whose elements are re-normalized to length $\sqrt{n}$.
Random time-frequency shift of the Alltop sequence \eqref{eq:mub} is a concrete example for such a sampling procedure, provided that the dimension $n \geq 5$ is an (odd) prime.

The vector $\vct{a}_{mub}$ is chosen from a union of $n$ bases that are all mutually unbiased with respect to the standard basis, see Eq.~\eqref{eq:mutually_unbiased}. Together with super-normalization ($\| \vct{a} \|_{\ell_2} = \sqrt{n}$) this readily establishes incoherence: $\max_{1 \leq k \leq n} | \langle \vct{e}_k,\vct{a} \rangle|^2 = \frac{n}{n}=1$ with probability one.

Next, by assumption $\vct{a}_{mub}$ is chosen uniformly from a union of $n$ re-scaled orthonormal bases $\left\{\sqrt{n}\vct{b}_1^{(l)},\ldots,\sqrt{n}\vct{b}_n^{(l)} \right\}$ with $1 \leq l \leq n$. Therefore, for any $\vct{z} \in \mathbb{C}^n$
\begin{align*}
\mathbb{E} \left[| \langle \vct{a}_{mub},\vct{z} \rangle|^2 \right]
=& \frac{1}{n^2} \sum_{l=1}^n \sum_{i=1}^n | \sqrt{n} \langle \vct{b}_i^{(l)},\vct{z} \rangle |^2 \\
=& \frac{1}{n} \sum_{l=1}^n \| \vct{z} \|_{\ell_2}^2 = \| \vct{z} \|_{\ell_2}^2
\end{align*}
which establishes isotropy.

Finally, a maximal set of $(n+1)$ mutually unbiased bases -- including the standard basis which we denote by $\vct{b}_k^{(n+1)} = \vct{e}_k$ -- forms a 2-design according to Definition~\ref{def:design}. For any $\vct{z} \in \mathbb{C}^n$ this property ensures
\begin{align*}
\mathbb{E} \left[ | \langle \vct{a}_{mub},\vct{z} \rangle|^4 \right]
=&  \sum_{l=1}^{n+1} \sum_{i=1}^n |\langle \vct{b}_i^{(l)}, \vct{z} \rangle|^4 
- \sum_{k=1}^n \left| \langle \vct{e}_k,\vct{z} \rangle \right|^4\\
=& 2\| \vct{z} \|_{\ell_2}^4 - \| \vct{z} \|_{\ell_4}^4 \leq 2 \| \vct{z} \|_{\ell_2}^4
\end{align*}
which implies the 4th moment bound.
In summary, the random vector $\vct{a}_{mub} \in \mathbb{C}^n$ meets the requirements of Theorem~\ref{thm:derandomization}. 
Theorem~\ref{thm:mub} then readily follows form the implications of the nullspace property for $s$-sparse recovery.

\section{Extension to noisy measurements} \label{sec:stable}

The nullspace property may be generalized to address two imperfections in $s$-sparse recovery simultaneously: (i) the vector $\vct{x} \in \mathbb{C}^d$ may only be approximately sparse in the sense that it is well-approximated by  a $s$-sparse vector, (ii) the measurements may be corrupted by additive noise: $\vct{y} = \mtx{A} \vct{x} +\vct{s}$ with $\vct{s} \in \mathbb{C}^m$.

To state this generalization, we need some additional notation. For $\vct{z} \in \mathbb{C}^n$ and $1 \leq s \leq n$, let $\vct{z}_s \in \mathbb{C}^n$ be the vector that only contains the $s$ largest entries in modulus. All other entries are set to zero. Likewise, we write $\vct{z}_{\bar{s}} = \vct{z} - \vct{z}_s$ to denote the remainder. In particular, $\sigma_s (\vct{z}) = \| \vct{z}_{\bar{s}} \|_{\ell_1}$.
A $m \times n$ matrix $\mtx{A}$ obeys the \emph{robust nullspace property} of order $s$ with parameters $\rho \in (0,1)$ and $\tau >0$ if
\begin{equation*}
\| \vct{z}_s \|_{\ell_2} \leq \frac{\rho}{\sqrt{s}} \| \vct{z}_{\bar{s}}\|_{\ell_1} + \tau \| \mtx{A} \vct{z} \|_{\ell_2}\quad \textrm{for all} \quad \vct{z} \in \mathbb{S}^{n-1}
\end{equation*}
see e.g.\ \cite[Definition~4.21]{foucart_mathematical_2013}. 
This extension of the nullspace property is closely related to stable $s$-sparse recovery from noisy measurements
via basis pursuit denoising:
\begin{align}
\textrm{minimize} & \quad \| \vct{z} \|_{\ell_1} \label{eq:bpdn} \\
\textrm{subject to} & \quad \| \mtx{A}\vct{z} - \vct{y} \|_{\ell_2} \leq \eta. \nonumber
\end{align}
Here, $\eta >0$ denotes an upper bound on the strength of the noise corruption: $\| \vct{s} \|_{\ell_2} \leq \eta$.
Indeed, \cite[Theorem~4.22]{foucart_mathematical_2013} draws the following connection: suppose that $\mtx{A}$ obeys the robust nullspace property with parameters $\rho,\tau$. Then, the solution $\vct{z}^\sharp \in \mathbb{C}^n$ to \eqref{eq:bpdn} is guaranteed to obey
\begin{align}
\| \vct{z}^\sharp - \vct{x} \|_{\ell_2} \leq \frac{D_1}{\sqrt{s}} \sigma_s (\vct{x}) + D_2 \eta, \label{eq:noisy_recovery}
\end{align}
where $D_1 = (1+\rho)^2/(1-\rho)$ and $D_2 = (3+\rho)\tau/(1-\rho)$. 
The first term on the r.h.s.\ vanishes if $\vct{x}$ is exactly $s$-sparse and remains small if $\vct{x}$ is well approximated by a $s$-sparse vector.
The second term scales linearly in the noise bound $\eta \geq \| \vct{s} \|_{\ell_2}$ and vanishes in the absence of any noise corruption. 

In the previous section, we have established the classical nullspace property for measurements that are chosen independently from a vector distribution that is isotropic, incoherent and obeys a bound on the 4th moments. 
This argument may readily be extended to establish the robust nullspace property with relatively little extra effort.
To this end, define the set
\begin{equation*}
T_{\rho,s} = \left\{\vct{z} \in\mathbb{S}^{n-1}:\; \| \vct{z}_s \|_{\ell_2} >\frac{\rho}{\sqrt{s}} \| \vct{z}_{\bar{s}}\|_{\ell_1} \right\} \subseteq \mathbb{S}^{n-1}.
\end{equation*}
A moment of thought reveals that the matrix $\mtx{A}$ obeys the robust nullspace property with parameters $\rho,\tau$ if
\begin{equation}
\inf_{\vct{z} \in T_{\rho,s}} \| \mtx{A} \vct{z} \|_{\ell_2} \geq \frac{1}{\tau}. \label{eq:nsp_reformulation}
\end{equation}
What is more, the following inclusion formula is also valid:
\begin{equation*}
T_{\rho,s} \subset \frac{3}{\rho} \mathrm{conv} \left( \Sigma_n^s \right),
\end{equation*}
see \cite[Lemma~3]{kabanava_analysis_2015} and \cite[Lemma~4.5]{rudelson_sparse_2008}. This ensures that the bounds on the parameters in Mendelson's small ball method generalize in a rather straightforward fashion. 
Isotropy, incoherence and the 4th moment bound ensure
\begin{align*}
Q_{2\xi} (\vct{a},T_{\rho,s}) \geq &\frac{(1-2\xi^2)^2}{C_4}, \\
 W_m (\vct{a},T_{\rho,s}) \leq & \frac{12}{\rho} \sqrt{2s \log (2n)}.
\end{align*}
Now, suppose that $\vct{A}$ subsumes $m \geq C \rho^{-2} s \log (2n)$ independent copies of the random vector $\vct{a} \in \mathbb{C}^n$, where $C>0$ is sufficiently large. Then, Theorem~\ref{thm:mendelson_complex} readily asserts
\begin{equation}
\inf_{\vct{z} \in T_{\rho,s}} \| \mtx{A} \vct{z} \|_{\ell_2} \geq \frac{c}{\rho} \sqrt{m} \label{eq:stable_lower_bound}
\end{equation}
with probability at least $1-2\mathrm{e}^{-\tilde{c}m}$. 
Previously, we employed Mendelson's small ball method to simply assert that a similar infimum is strictly positive. Eq.~\eqref{eq:stable_lower_bound} provides a strictly positive lower bound with comparable effort. 
Comparing this relation to Eq.~\eqref{eq:nsp_reformulation} 
highlights that this is enough to establish the robust nullspace property with parameters $\rho$ and $\tau = \frac{\rho}{c \sqrt{m}}$ with high probability. In turn, a stable generalization of the main recovery guarantee follows from Eq.~\eqref{eq:noisy_recovery}.

\begin{theorem}
Fix $\rho \in (0,1)$ and $s \in \mathbb{N}$.
Suppose that we sample $m \geq C \rho^{-2} s \log (n)$ independent copies of an isotropic, incoherent random vector $\vct{a} \in \mathbb{C}^n$ that also obeys the 4th moment bound. 
Then, with probability at least $1-2\mathrm{e}^{-\tilde{c}m}$, the resulting measurement matrix $\mtx{A}$ allows for stable, uniform recovery of (approximately) $s$-sparse vectors. 
More precisely, the solution $\vct{z}^\sharp$ to \eqref{eq:bpdn} is guaranteed to obey
\begin{align*}
\| \vct{x} - \vct{z}^\sharp \|_{\ell_2} \leq \frac{D_1}{\sqrt{s}} \sigma_s (\vct{x}) + D_2 \frac{\eta}{\sqrt{m}},
\end{align*}
where $D_1,D_2 >0$ depend only on $\rho$.
\end{theorem}

\section{Numerical experiments} \label{sec:numerics}

In this part we demonstrate the performance which can be achieved with
our proposed derandomized constructions and we compare this to
generic measurement matrices (Gaussian, signed Bernoulli). However, since
the orthogonal array construction is more involved we first provide
additional details relevant for numerical experiments.

\subsection{Details on orthogonal arrays} An orthogonal array
$\OA(\OAindex\OAlevels^\OAstrength,\OAfactors,\OAlevels,\OAstrength)$
of strength $\OAstrength$, with $\OAfactors$ factors and $\OAlevels$
levels is an $\OAindex\OAlevels^\OAstrength\times \OAfactors$ array of
$\OAlevels$ different symbols such that in any $\OAstrength$ columns
every ordered $\OAlevels^\OAstrength$-tuple occurs in exactly
$\OAindex$ rows. Arrays with $\OAindex=1$ are called simple.  A
comprehensive treatment can be found in the book
\cite{Hedayat:OAbook}. Known arrays are listed in several
libraries\footnote{for example \url{http://neilsloane.com/oadir/} or
  \url{http://pietereendebak.nl/oapage/}}.  Often the symbol alphabet
is not relevant, but we use the set
$\OAsymbols_{\OAlevels}=\{0,\dots, \OAlevels-1\}$ for concreteness.  Such arrays can be represented as a matrix in
$\OAsymbols_{\OAlevels}^{\OAindex\OAlevels^\OAstrength\times\OAfactors}$.
For $\OAlevels=q^p$ with $q$ prime the simple orthogonal array
$\OA(\OAlevels^\OAstrength,\OAfactors,\OAlevels,\OAstrength)$ is {\em linear} if
the $q^{pt}$ rows of the matrix form a vector space over
$\mathbb{F}_q$.
The runs of an orthogonal array (the rows of the corresponding matrix)
can also be interpreted as codewords of a code and vice
versa. The array is linear if and only if the corresponding code is linear
\cite[Chapter 4]{Hedayat:OAbook}.
This relationship allows to employ
classical code constructions to construct orthogonal arrays.

\subsection{Counting bits}
In this work we propose to generate $m\times n$ sampling matrices
$\mtx{A}$ by selecting $m\leq \OAruns=\OAindex\OAlevels^\OAstrength$
rows at random from an orthogonal array
$\OA(\OAindex\OAlevels^4,n,\OAlevels,4)$, eventually
removing the bias (substracting $(\OAlevels-1)/2$ per component) and
scale appropriately. Intuitively, $m\log_2(\OAruns)$ bits are
then required to specify such a matrix $\mtx{A}$.  For $\OAstrength=4$
and $k=n$, a classical lower bound due to Rao \cite{Rao:1947} demands
\begin{equation}
  M=\OAindex\OAlevels^4\geq1+\OAfactors+\binom{\OAfactors}{2}=\Omega(\OAfactors^2).
\end{equation}
Arrays that saturate this bound are called tight (or complete).
In summary, an order of $s\log^2(n)$ bits are required to
sample a $m \times n$ matrix $\mtx{A}$ with $m \geq C s \log (n)$ rows
according to this procedure.

\subsection{Strength-$4$ Constructions}

For compressed
sensing applications we want arrays with large number of factors $\OAfactors$ since this
corresponds to the ambient dimension $n=k$ of the sparse vectors to recover. On the other
hand the run size $\OAruns$ should scale ``moderately'' to describe
the random matrices only with few bits.
Most constructions use an existing orthogonal array as a {\em seed}
to construct larger arrays. Known binary arrays of strength $4$ are for
example the simple array $\OA(16,5,2,4)$, or
$\OA(80,6,2,4)$.
Ref.~\cite{Pat2012} proposes an algorithm
that uses a linear orthogonal array $\OA(N,\OAfactors,\OAlevels,\OAstrength)$
as a seed to construct a linear orthogonal array
$\OA(N^2,\OAfactors^2+2\OAfactors,\OAlevels,\OAstrength)$. This procedure may then be iterated.

\subsection{Numerical results for orthogonal arrays:}
\begin{figure*}[h]
\begin{minipage}{.5\textwidth}
\centering
  \includegraphics[width=\linewidth]{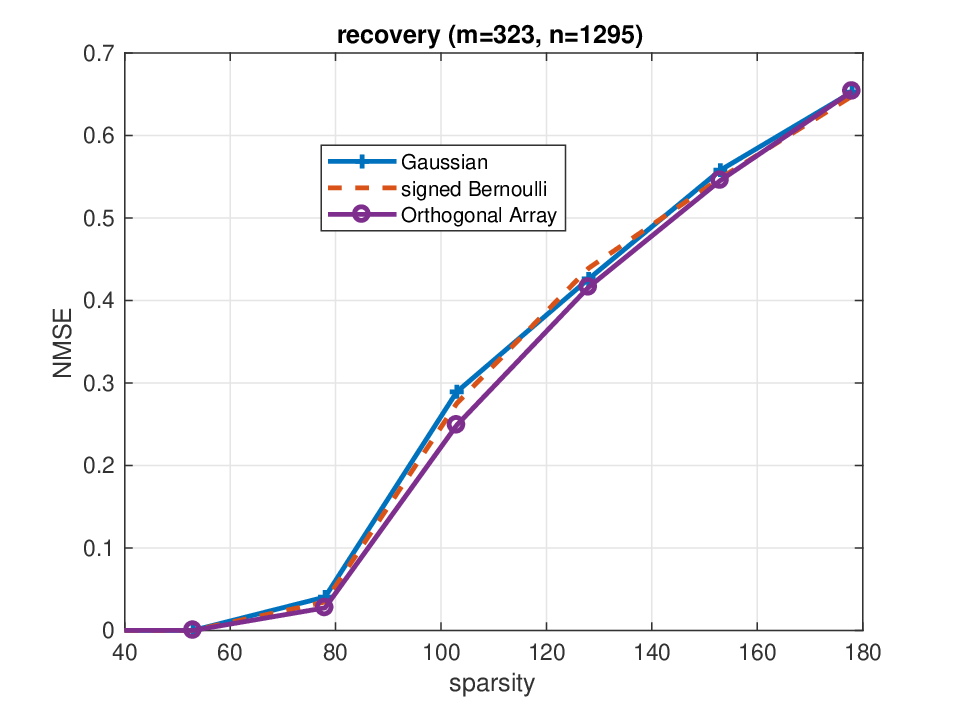}
\end{minipage}
\begin{minipage}{.5\textwidth}
\centering
  \includegraphics[width=\linewidth]{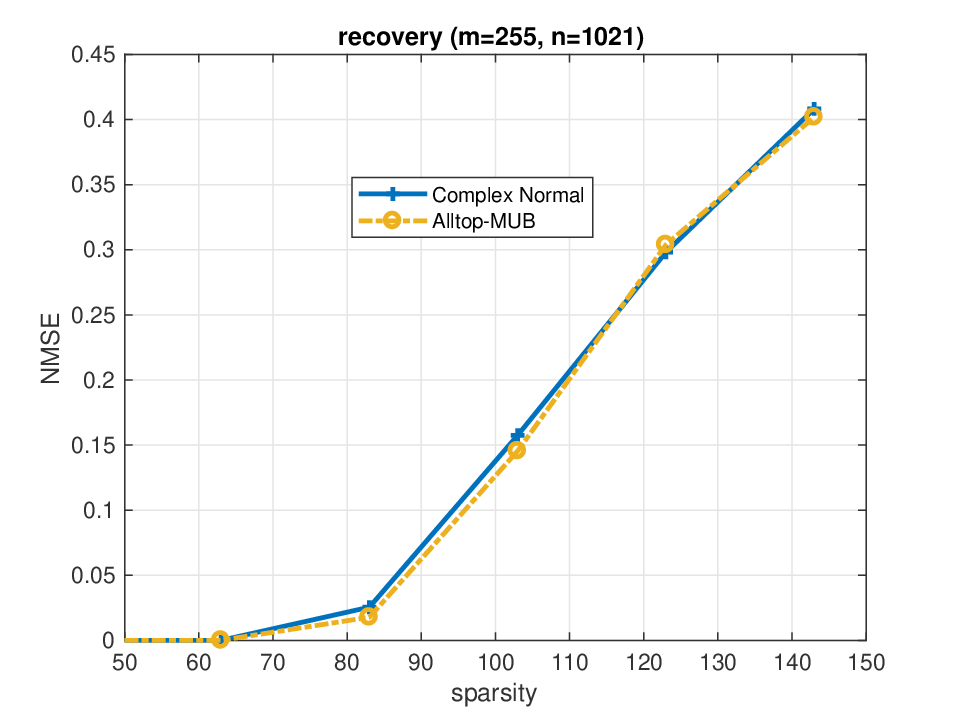}
\end{minipage}
  \caption{\emph{Left:} Performance of basis pursuit for $m=323$ and $n=1295$ (real-valued signals) from random orthogonal array measurements and their generic counterpart (signed Bernoulli). \newline
\emph{Right:} 
Performance of basis pursuit for $m=255$ and $n=1021$ (complex-valued signals) from random time-frequency shifted Alltop sequences and their generic counterpart (standard complex Gaussian vectors).
  }
  \label{fig:plots}
\end{figure*}

Figure \ref{fig:plots} summarizes the empirical
performance of basis pursuit \eqref{eq:l1minimization} from independent orthogonal array measurements. We consider real-valued signals and quantify the performance in terms of the normalized
$\ell_2$-recovery error (NMSE).
To construct the orthogonal array,
algorithm \cite{Pat2012} is applied twice
$\OA(16,5,2,4)
\rightarrow\OA(256,35,2,4)\rightarrow\OA(65536,1295,2,4)$.
The $323$ rows are uniformly sampled from this array, i.e. 
the sampling matrix $\mtx{A}$ has $\pm1$ entries (mapping
$\{0,1\}\rightarrow\{\pm 1\}$) and size
$323\times 1295$. Note that, in the case of non-negative sparse
vectors, the corresponding 0/1-matrices may be used instead
to recover with non-negative least-squares
\cite{kueng_robust_2018}.  The sparsity of the unknown vector has been
varied between $1\dots180$. For each sparsity many experiments are
performed to compute NMSE. In each run, the support of the
unknown vector has been chosen uniformly at random and the values are independent instances of a standard Gaussian random variable. 
For comparison, we have also included the corresponding performances of a generic sampling matrix (signed Bernoulli) of the same size.
Numerically, the partially derandomized orthogonal array construction achieves essentially the same performance as its generic counterpart.

\subsection{Numerical results for the Alltop design}
Figure~\ref{fig:plots} shows the NMSE achieved for measurement
matrices based on subsampling from an Alltop-design \eqref{eq:mub}. The data is obtained in the same way as above but the sparse vectors are
generated as iid. complex-normal distributed on the support.
For comparison the results for a (complex) standard Gaussian sampling matrix are included as well.
Again, the performance of random Alltop-design measurements essentially matches its generic (Gaussian) counterpart.

\section{Additional proofs}

\subsection{Proof of Lemma~\ref{lem:W}} \label{sub:W}

The inclusion $T_s \subset 2 \mathrm{conv} (\Sigma_n^s)$ remains valid in the complex case. Moreover, every $\vct{z} \in \mathrm{conv}(\Sigma_n^s)$ necessarily obeys
\begin{equation*}
\max_{\vct{z} \in \mathrm{conv}(\Sigma_n^s)} \| \vct{z} \|_{\ell_1} \leq \max_{\vct{z} \in \Sigma_n^s} \| \vct{z} \|_{\ell_1} = \sqrt{s},
\end{equation*}
because the maximum value of a convex function over a convex set is achieved at the boundary. Hoelder's inequality therefore implies
\begin{align}
W_m (\vct{a},T_s) =& \mathbb{E} \sup_{\vct{z} \in T_s} | \langle \vct{z},\vct{h} \rangle \leq 2 \mathbb{E} \sup_{\vct{z} \in \mathrm{conv}(\Sigma_n^s)} \| \vct{z} \|_{\ell_1} \| \vct{h} \|_{\ell_\infty} \nonumber \\
\leq & 2 \sqrt{s} \mathbb{E} \| \vct{h} \|_{\ell_\infty}, \label{eq:Waux1}
\end{align}
where $\vct{h} = \frac{1}{\sqrt{m}} \sum_{i=1}^m \epsilon_i \vct{a}_i \in \mathbb{C}^n$.
Moreover,
\begin{align*}
\mathbb{E} \| \vct{h} \|_{\ell_\infty} 
=& \mathbb{E} \max_{1 \leq k \leq n} | \langle \vct{e}_k, \vct{h} \rangle | \\
\leq & \mathbb{E} \max_{1 \leq k \leq n} \left| \mathrm{Re} (\langle \vct{e}_k, \vct{h} \rangle) \right| + \mathbb{E} \max_{1 \leq k \leq n} | \mathrm{Im} (\langle \vct{e}_k, \vct{h} \rangle |
\end{align*}
and we may bound both expressions on the r.h.s.\ independently. 
For the first term, fix $\theta>0$ and use Jensen's inequality (the logarithm is a concave function) to obtain 
\begin{align*}
&\mathbb{E} \max_{1 \leq k \leq n} | \mathrm{Re}( \langle \vct{e}_k, \vct{h} \rangle )|
= \mathbb{E} \max_{1 \leq k \leq n} \max_{\sigma = \pm} \sigma \mathrm{Re} (\langle \vct{e}_k, \vct{h} \rangle ) \\
\leq & \frac{1}{\theta} \log \left( \mathbb{E} \exp \left( \max_{1 \leq k \leq n} \max_{\sigma = \pm} \theta \sigma \mathrm{Re} \left( \langle \vct{e}_k, \vct{h} \rangle \right) \right) \right).
\end{align*}
Monotonicity and non-negativity of the exponential function then imply
\begin{align*}
&  \mathbb{E} \exp \left( \max_{1 \leq k \leq n} \max_{\sigma = \pm} \theta \sigma \mathrm{Re} \left( \langle \vct{e}_k, \vct{h} \rangle \right) \right) \\
\leq  &\sum_{k=1}^n \sum_{\sigma = \pm} \mathbb{E} \exp \left( \theta \sigma \mathrm{Re} \left( \langle \vct{e}_k, \vct{h} \rangle \right) \right) \\
=& \sum_{k=1}^n \sum_{\sigma = \pm} \prod_{i=1}^m \mathbb{E} \exp \left( \frac{ \theta \sigma}{\sqrt{m}} \epsilon_i \mathrm{Re} \left( \langle \vct{e}_k, \vct{a}_i \rangle \right) \right),
\end{align*}
where we have also used that all $\epsilon_i$'s and $\vct{a}_i$'s are independent. 
The remaining moment generating functions can be bounded individually. Fix $1 \leq k \leq n$, $\sigma \in \left\{ \pm 1 \right\}$ and $1 \leq i \leq m$ and exploit the Rademacher randomness to infer
\begin{align*}
& \mathbb{E} \exp \left( \frac{\theta \sigma}{\sqrt{m}} \epsilon_i \mathrm{Re} \left( \langle \vct{e}_k, \vct{a}_i \rangle \right) \right)
= \mathbb{E} \cosh \left( \frac{\theta \sigma}{\sqrt{m}} \mathrm{Re} \left( \langle \vct{e}_k, \vct{a}_i \rangle \right) \right) \\
\leq &\mathbb{E} \exp \left( \frac{\theta^2 \sigma^2}{2m} \mathrm{Re} \left( \langle \vct{e}_k, \vct{a}_i \rangle \right)^2 \right) 
= \mathbb{E} \exp \left( \frac{\theta^2}{2m} \mathrm{Re} \left( \langle \vct{e}_k, \vct{a}_i \rangle \right)^2 \right),
\end{align*}
because $\sigma^2 =1$. Incoherence moreover ensures $(\mathrm{Re}(\langle \vct{e}_k, \vct{a}_i \rangle)^2 \leq| \langle \vct{e}_k, \vct{a}_i \rangle|^2 \leq 1$. This ensures that the remaining expectation value is upper-bounded by $\exp \left(\frac{\theta^2}{2m} \right)$. Inserting these individual bounds into the expression above yields
\begin{align*}
& \mathbb{E} \max_{1 \leq k \leq n} \left| \mathrm{Re}(\langle \vct{e}_k,\vct{h} \rangle ) \right| \\
 \leq &\frac{1}{\theta} \log \left( \sum_{k=1}^n \sum_{\sigma = \pm} \prod_{i=1}^m \mathbb{E} \exp \left( \frac{\theta \sigma}{\sqrt{m}} \epsilon_i \mathrm{Re} \left( \langle \vct{e}_k,\vct{a}_i \rangle \right) \right) \right) \\
\leq & \frac{1}{\theta} \log \left( \sum_{k=1}^n \sum_{\sigma=\pm} \prod_{i=1}^m \exp \left( \frac{\theta^2}{2m} \right) \right) \\
=& \frac{1}{\theta} \log \left( 2n \exp \left( \frac{\theta^2}{2} \right) \right) 
= \frac{ \log (2n)}{\theta} + \frac{\theta}{2}
\end{align*}
for any $0 <\theta \leq \sqrt{2m}$. Choosing $\theta = \sqrt{2\log(2n)}$ is feasible and minimizes this upper bound. A completely analogous bound can be derived for the expected maximum absolute value of the imaginary part. Combining both yields
\begin{align*}
\mathbb{E} \| \vct{h} \|_{\ell_\infty}
\leq  \sqrt{ 2\log (2n)} +  \sqrt{2\log (2n)} = 2 \sqrt{2 \log (2n)}
\end{align*}
and inserting this bound into Eq.~\eqref{eq:Waux1} ensures
\begin{equation*}
W_m (\vct{a},T_s) \leq 4 \sqrt{2s \log (2n)}.
\end{equation*}

\subsection{Proof of Theorem~\ref{thm:mendelson_complex}} \label{sub:mendelson_proof}

The proof is based on rather straightforward modifications of Tropp's proof for Mendelson's small ball method \cite{tropp_convex_2015}.
Let $\vct{a} \in \mathbb{C}^n$ be a complex-valued random vector. Suppose that $\vct{a}_1,\ldots,\vct{a}_m \in \mathbb{C}^n$ are independent copies of $\vct{a}$ and let $\mtx{A}$ be the $m \times n$ matrix whose $m$ rows correspond to these vectors.
The goal is to obtain a lower bound on
$
\inf_{\vct{z} \in E} \| \mtx{A} \vct{z} \|_{\ell_2},
$
where $E \subset \mathbb{C}^n$ is an arbitrary, but fixed, set.
First, note that $\ell_1$ and $\ell_2$ norms on $\mathbb{R}^{2m}$ are related via $\| \vct{v} \|_{\ell_2} \geq (2m)^{-1} \| \vct{v} \|_{\ell_1}$.
For fixed $\vct{z} \in E$ this ensures
\begin{align*}
 \| \mtx{A} \vct{z} \|_{\ell_2} = &\sqrt{ \sum_{i=1}^m | \langle \vct{a}_i, \vct{z} \rangle|^2} \\
=& \sqrt{ \sum_{i=1}^m\mathrm{Re} \left( \langle \vct{a}_i, \vct{z} \rangle \right)^2 + \sum_{i=1}^m \mathrm{Im} \left( \langle \vct{a}_i, \vct{z} \rangle \right)^2} \\
\geq & \frac{1}{\sqrt{2m}} \left( \sum_{i=1}^m | \mathrm{Re}(\langle \vct{a}_i, \vct{z} \rangle) | + \sum_{i=1}^m | \mathrm{Im} (\langle \vct{a}_i, \vct{z} \rangle) | \right) \\
=& \frac{1}{\sqrt{2m}} \sum_{i=1}^m \left(  | \mathrm{Re}(\langle \vct{a}_i, \vct{z} \rangle) |+ | \mathrm{Im}(\langle \vct{a}_i, \vct{z} \rangle) | \right).
\end{align*}
Next, we fix $\xi >0$ arbitrary and introduce the indicator function $\mathbb{I} \left\{ x \geq \xi \right\}$ which obeys $x \geq \xi \mathbb{I} \left\{ x \geq \xi \right\}$ for all $x \geq 0$. Consequently, $\| \mtx{A} \vct{z} \|_{\ell_2}$ is upper-bounded by
\begin{align}
 \frac{\xi}{\sqrt{2m}} \sum_{i=1}^m \left( \mathbb{I} \left\{ | \mathrm{Re}(\langle \vct{a}_i, \vct{z} \rangle) |\geq \xi \right\} + \mathbb{I} \left\{ | \mathrm{Im}(\langle \vct{a}_i, \vct{z} \rangle) | \geq \xi \right\} \right). \label{eq:complex_mendelson_aux1}
\end{align}
Also, note that the expectation value of each summand obeys
\begin{align*}
& \mathbb{E} \left[  \mathbb{I} \left\{ | \mathrm{Re}(\langle \vct{a}_i, \vct{z} \rangle) |\geq \xi \right\} \right] + \mathbb{E} \left[   \mathbb{I} \left\{ | \mathrm{Im}(\langle \vct{a}_i, \vct{z} \rangle) |\geq \xi \right\}\right] \\
=& \mathrm{Pr} \left[   | \mathrm{Re}(\langle \vct{a}_i, \vct{z} \rangle) | \geq \xi \right] + \mathrm{Pr} \left[ | \mathrm{Im}(\langle \vct{a}_i, \vct{z} \rangle) | \geq \xi \right] \\
\geq & \mathrm{Pr} \left[ | \mathrm{Re}(\langle \vct{a}_i, \vct{z} \rangle) | \geq \xi \; \vee \; | \mathrm{Im}(\langle \vct{a}_i, \vct{z} \rangle) | \geq \xi \right] \\
\geq & \mathrm{Pr} \left[ | \langle \vct{a}_i, \vct{z} \rangle| \geq \sqrt{2} \xi \right],
\end{align*}
according to the union bound. 
The last line follows from the following observation. Let $z = a+ib$ be a complex number. Then, $|z| = \sqrt{a^2+b^2} \geq \sqrt{2} \xi$ necessarily implies either $|a| \geq \xi$, or $|b| \geq \xi$ (or both).
Now, define 
\begin{equation*}
Q_{2\xi} (\vct{z}) = \mathrm{Pr} \left[   | \mathrm{Re}(\langle \vct{a}_i, \vct{z} \rangle) | \geq 2\xi \right] + \mathrm{Pr} \left[ | \mathrm{Im}(\langle \vct{a}_i, \vct{z} \rangle) | \geq 2\xi \right]
\end{equation*}
and note that the estimate from above ensures
\begin{align}
\inf_{\vct{z} \in E}Q_{2\xi} (\vct{z}) \geq &\inf_{\vct{z} \in E} \mathrm{Pr} \left[ | \langle \vct{a},\vct{z} \rangle| \geq 2^{3/2} \xi \right] \nonumber \\
=& Q_{2^{3/2} \xi} (\vct{a},E).
\label{eq:complex_mendelson_aux2}
\end{align}
Adding and subtracting $\xi (m/2)^{1/2} Q_{2\xi}(\vct{z})$ to Eq.~\eqref{eq:complex_mendelson_aux1} and taking the infimum yields

\begin{align}
& \inf_{\vct{z} \in E} \| \mtx{A} \vct{z} \|_{\ell_2} \nonumber \\
\geq &\inf_{\vct{z} \in E}
\bigg( \xi \sqrt{\frac{m}{2}} Q_{2\xi} (\vct{z}) - \xi \sqrt{\frac{m}{2}}Q_{2\xi} (\vct{z})  \nonumber \\
+&  \frac{\xi}{\sqrt{2m}} \sum_{i=1}^m \left( \mathbb{I} \left\{ | \mathrm{Re}(\langle \vct{a}_i, \vct{z} \rangle) |\geq \xi \right\} + \mathbb{I} \left\{ | \mathrm{Im}(\langle \vct{a}_i, \vct{z} \rangle) | \geq \xi \right\} \right) \bigg) \nonumber \\
\geq & \xi \sqrt{\frac{m}{2}} Q_{2^{3/2} \xi}(\vct{a},E) 
 - \frac{\xi}{\sqrt{2m}} \sup_{\vct{z} \in E} \bigg( m Q_{2\xi}(\vct{z})  \nonumber \\
-&  \sum_{i=1}^m \!\left( \mathbb{I} \left\{ | \mathrm{Re}(\langle \vct{a}_i, \vct{z} \rangle) |\geq \xi \right\}\!+\!\mathbb{I} \left\{ | \mathrm{Im}(\langle \vct{a}_i, \vct{z} \rangle) | \geq \xi \right\} \right) \bigg). \label{eq:complex_mendelson_aux3}
\end{align}
Here we have applied Eq.~\eqref{eq:complex_mendelson_aux2} to the first term. Since $Q_{2\xi}(\vct{z})$ features both a real and imaginary part and we can split up the remaining supremum accordingly. 
The suprema over real and complex parts individually correspond to
\begin{align*}
\sup_{\vct{z} \in E} \sum_{i=1}^m 
\left( \mathrm{Pr} \left[ |\mathrm{Re}(\langle \vct{a}_i,\vct{z} \rangle)| \geq 2 \xi \right] - \mathbb{I} \left\{  |\mathrm{Re}(\langle \vct{a}_i,\vct{z} \rangle)| \geq \xi \right\} \right), \\
\sup_{\vct{z} \in E} \sum_{i=1}^m \left( \mathrm{Pr} \left[ |\mathrm{Im}(\langle \vct{a}_i,\vct{z} \rangle)| \geq 2 \xi \right] - \mathbb{I} \left\{  |\mathrm{Im}(\langle \vct{a}_i,\vct{z} \rangle)| \geq \xi \right\} \right) 
\end{align*}
and we denote them by $R(E,\vct{a})$ and $I(E,\vct{a})$, respectively.
The vectors $\vct{a}_1,\ldots,\vct{a}_m$ are independent copies of $\vct{a} \in \mathbb{C}^n$.
The bounded difference inequality \cite[Section~6.1]{boucheron_concentration_2013} asserts that both expressions concentrate around their expectation. More precisely, for any $t >0$ 
\begin{align*}
\mathrm{Pr} \left[ R(E,\vct{a}) \geq  \mathbb{E} R(E,\vct{a})+t \sqrt{m} \right] \leq & \mathrm{e}^{-t^2/2}, \\
\mathrm{Pr} \left[ I(E,\vct{a}) \geq \mathbb{E} I(E,\vct{a})+t \sqrt{m} \right] \leq & \mathrm{e}^{-t^2/2}.
\end{align*}
Therefore, the union bound grants a transition from $R(E,\vct{a})+I (E,\vct{a})$ to $\mathbb{E} R(E,\vct{a}) + \mathbb{E} I (E,\vct{a}) + 2 \sqrt{m}t$
with probability at least $1-2\mathrm{e}^{-t^2/2}$.
These expectation values can be further simplified. Define the soft indicator function
\begin{equation*}
\psi_\xi (s) = 
\begin{cases}
0 & |s| \leq \xi, \\
(|s|-\xi)/\xi & \xi \leq |s| \leq 2\xi, \\
1 & |s| \geq 2 \xi
\end{cases}
\end{equation*}
which obeys $\mathbb{I} \left\{ |s| \geq 2 \xi \right\} \leq \psi_\xi (s) \leq \mathbb{I} \left\{ |s| \geq \xi \right\}$ for all $s \in \mathbb{R}$.
Moreover, $\xi \psi_\xi (s)$ is a contraction, i.e.\ a real-valued function with Lipschitz constant one that also obeys $\xi \psi_\xi (0) = 0$.
Rademacher symmetrization \cite[Lemma~8.4]{foucart_mathematical_2013} and the Rademacher comparison principle \cite[Eq.~(4.20)]{ledoux_probability_2013} yield
\begin{align*}
& \mathbb{E}\; R(E,\vct{a}) \\
=& \mathbb{E} \sup_{\vct{z} \in E} \sum_{i=1}^m \left( \mathbb{E} \mathbb{I} \left\{ |\mathrm{Re}(\vct{a}_i,\vct{z})| \geq 2 \xi \right\} - \mathbb{I} \left\{ | \mathrm{Re}(\langle \vct{a}_i,\vct{z} \rangle)| \geq \xi \right\} \right) \\
\leq & \mathbb{E} \sup_{\vct{z} \in E} \sum_{i=1}^m \left( \mathbb{E} \psi_\xi ( \mathrm{Re}(\langle \vct{a}_i,\vct{z} \rangle)) - \psi_\xi ( \mathrm{Re}(\langle \vct{a}_i,\vct{z} \rangle) \right) \\
\leq & 2 \mathbb{E} \sum_{i=1}^m \epsilon_i \psi_\xi ( \mathrm{Re}(\langle \vct{a}_i, \vct{z} \rangle ) 
\leq \frac{2}{\xi} \mathbb{E} \sup_{\vct{z} \in E} \sum_{i=1}^m \epsilon_i \mathrm{Re} (\langle \vct{a}_i,\vct{z} \rangle) \\
\leq & \frac{2\sqrt{m}}{\xi}\mathbb{E} \sup_{\vct{z} \in E} \left| \langle  \vct{z},\vct{h} \rangle \right|,
\end{align*}
where $\vct{h} = \frac{1}{\sqrt{m}} \sum_{i=1}^m \epsilon_i \vct{a}_i \in \mathbb{C}^n$. 
A completely analogous bound holds true for $\mathbb{E} I (E,\vct{a})$. 
Inserting both bounds into Eq.~\eqref{eq:complex_mendelson_aux3}
establishes
\begin{align*}
& \inf_{\vct{z} \in E} \| \mtx{A} \vct{z} \|_{\ell_2} \\
\geq &\xi \sqrt{\frac{m}{2}} Q_{2^{3/2} \xi} - \frac{\xi}{\sqrt{2m}} \left( \frac{4\sqrt{m}}{\xi} \mathbb{E} \sup_{\vct{z} \in E} | \langle \vct{z},\vct{h} \rangle| + 2 \sqrt{mt} \right) \\
=& \xi \sqrt{\frac{m}{2}} Q_{2^{3/2} \xi} - 2^{3/2} \mathbb{E} \sup_{\vct{z} \in E} | \langle \vct{z},\vct{h} \rangle| - \sqrt{2} \xi t
\end{align*}
with probability at least $1-2\mathrm{e}^{-t^2/2}$.
Setting $W_m (E,\vct{z}) = \mathbb{E} \sup_{\vct{z} \in E} | \langle \vct{z},\vct{h} \rangle|$ establishes the claim.

\paragraph*{Acknowledgements}

This work can be seen as a continuation of the research program that David Gross devised for RK's doctoral studies. 
PJ is supported by DFG grant JU 2795/3 and DAAD grant 57417688.
RK was in part supported by Joel A. Tropp under ONR Award No. N00014-17-12146
and also acknowledges funding provided by the Institute of Quantum Information
and Matter, an NSF Physics Frontiers Center (NSF Grant PHY-1733907).
DGM was partially supported by AFOSR FA9550-18-1-0107, NSF DMS 1829955, and the Simons Institute of the Theory of Computing.

  \bibliographystyle{IEEEtran}
  \bibliography{cs_arrays}

\end{document}